# DENSE MOLECULAR CLOUDS IN THE GALACTIC CENTER REGION
# I. HCN (J=1–0) DATA[1]

Chang Won Lee

Department of Earth Science, Pusan National University, Pusan 609-735, Korea

and

Department of Astronomy and Space Science, Chungnam National University, Korea

E-mail: cwl@astrophys.es.pusan.ac.kr, cwl@sarang.chungnam.ac.kr

[1]Submitted to the Astrophysical Journal Supplement Series

## ABSTRACT

The Galactic Center regions $-6° \leq l \leq 6°$ and $-0.8° \leq b \leq 0.87°$ are surveyed with HCN J=1−0 line (88.6 GHz) on a $4'$ grid with the 14 m telescope at the Taeduk Radio Astronomy Observatory. The data complement the previous $^{13}$CO, CS, and $H_2CO$ measurements (Bally *et al.* 1988, Zylka *et al.* 1992) but cover a wider area with a better or comparable resolution.

The asymmetry in emission found in the $^{13}$CO observations by Bally *et al.* (1988) is well confirmed in this study. Interestingly, the asymmetry in HCN emission is more pronounced here: more than 80% of the overall emission is found in the first quadrant ($l > 0$) and $V_{LSR} > 0$. Also confirmed is the existence of several emission components in the regions forbidden by the Galactic circular rotation.

The HCN maps show clear absence of both more outer galactic emissions and the 3 kpc arm component, which are both visible in the $^{13}$CO map. This implies that the HCN emission lines are not contaminated by foreground or background gas components. Therefore HCN data are particularly useful in investigating the structures of molecular clouds in the Galactic Center. The $\Delta V$s (FWHMs) of the most of the detected lines are found to be very large, *i.e.*, about 15 ∼55 km s$^{-1}$ with the mode at $\Delta V \approx 25$ km s$^{-1}$. The typical value of $\Delta V \approx 25$ km s$^{-1}$ may imply that clouds with mass of an order of $10^6$ $M_\odot$ are common in the GC region. Another trend found is that the maximum $\Delta V$ decreases with $l$, which may be partly due to the overlapping effects of clouds.

## 1. INTRODUCTION

The Galactic Center (GC) region is distinct in many aspects. Matter concentration is high and there is energetic activity. Due to the influence of such peculiarities, molecular clouds in the GC region show unique physical properties and kinematics. During the last decade, numerous surveys in different molecular lines have been made in order to understand the global properties of the GC molecular clouds and their implications. Those have provided important information on the gravitational potential in the GC region (Binney *et al.* 1991) as well as physical properties of GC molecular clouds themselves. Our knowledge on the physics of the molecular clouds in the GC on a few hundred parsec scale is still limited, however, due to a rather poor space coverage and grid resolution of the previous surveys. Table 1 lists previous molecular line surveys toward the GC made with area coverage larger than $3^\circ \times 0.8^\circ$ and grid resolution better than $12' \times 12'$.

As tracers of molecular gas distribution, it is the CO isotopes that have been used most commonly, among which $^{12}$CO molecule has been most widely used for surveys. Since J=1−0 line (115 GHz) of $^{12}$CO is easily saturated in directions toward the GC and is, moreover, is severely contaminated by diffuse outer galaxy gas emission, it can not be a good tracer in the GC region. Instead, much rarer and thus optically thin molecular lines such as C$^{18}$O J=1−0 (109 GHz) could be a good tracer for a column density, except that the brightness temperature is normally too low for a large scale survey in the GC regions. Bally *et al.* (1987, 1988) have chosen $^{13}$CO J=1−0 (110 GHz) line as a compromise between the brighter $^{12}$CO J=1−0 line and the fainter C$^{18}$O J=1−0 line to survey the region of $-5^\circ \leq l \leq 5^\circ$ and $-0.4^\circ \leq b \leq 0.4^\circ$. Their study provided important information on the GC molecular clouds. However, saturation toward some regions is apparent in $^{13}$CO J=1−0 line, as is contamination by diffuse outer Galaxy disk gas components near $V_{LSR} \approx 0$ km s$^{-1}$, since the $^{13}$CO J=1−0 line can be excited into emission in relatively low density regions of n≈ 500 $cm^{-3}$. One of the distinct properties of the GC molecular clouds is that they are denser than a typical cloud in the Galactic plane by two orders of magnitude in order to remain bound in the enhanced tidal force in the inner Galaxy (Stark *et al.* 1989, Combes 1991). To observe the clouds in the GC, it is therefore essential to choose a line that is optically thin and traces a high density. Candidates for the line include CS J=2-1 (98 GHz) and $H_2CO$ $1_{10} - 1_{11}$ (4.83 GHz) and these have been used previously (CS: Bally *et al.* 1987, 1988; $H_2CO$: Zylka *et al.* 1992), but the surveys were limited to a rather small region.

In this paper the HCN J=1−0 line is chosen as an alternate tracer and the results of the survey toward the GC regions are presented. The rotational transition J=1−0 line (88.6 GHz) of HCN is an optically thin ($\tau \leq 0.3$), high density tracer (J=0−1 excitation requires n$\geq 10^5$ $cm^{-3}$ [Turner 1974]), which is bright enough to survey wide regions in the GC. Moreover, the Earth's atmosphere is more transparent at J=1−0 line of HCN than at the corresponding lines of $^{12}CO$ and $^{13}CO$. This is a strong advantage for low elevation observations of sources such as the present GC molecular clouds. This HCN (J=1−0) survey extends to a wider region ($-6° \leq l \leq 6°$, $-0.8° \leq b \leq 0.87°$) with a better or comparable grid resolution (4′) than earlier $^{13}CO$, CS, and $H_2CO$ measurements [1].

In §2 the observations and the data reduction processes are described. The global distribution, kinematics and some properties of the GC molecular clouds are described with various maps in §3. A summary of the results is given in the final section. Implications of the kinematics of GC molecular clouds will be discussed in forthcoming paper.

## 2. OBSERVATIONS AND DATA REDUCTIONS

[1]There is another recent HCN J=1-0 survey toward the Galactic Center region by Jackson *et al.* (1996) which is sampled on 1′ grid over $-2° \leq l \leq 2°$ and $-0.2° \leq b \leq 0.2°$.

A survey of the HCN J=1−0 transition ($\nu$ = 88.631847 GHz) was carried out in the region of $-6^\circ \leq l \leq 6^\circ$ and $-0.8^\circ \leq b \leq 0.87^\circ$ with the 14 m telescope of the Taeduk Radio Astronomy Observatory (TRAO) in Korea between October 1993 and May 1994. All of the data were recorded on a regular $4'$ grid, except those in the regions of weak emission which were sampled on a $8'$ grid. The total number of observed points was $\sim$ 1650. These points are shown in Figure 1. Spectra that are typical of the eight bright regions are shown in Figure 2. The HPBW and the main beam efficiency of the telescope at 88.6 GHz were $61''$ and 0.41, respectively (Park *et al.* 1994). The receiver was a cryogenic Schottky diode mixer type and the filter bank had 256 channels of 1 MHz band width each, which covered a total velocity range of $\sim$890 km s$^{-1}$.

Spectral line intensities were calibrated and corrected for atmospheric losses using the chopper wheel method to obtain the antenna temperature $T_A^*$. System temperatures during the course of the observations ranged from 500 to 900 K, typically near 600 K. Most of the data were taken with a position switching mode. Reference positions were chosen on the emission free regions about 1° above or below the source in Galactic latitude. Integration on a source varied between 3$\sim$5 minutes, which resulted in rms noises between 0.05 K $\sim$ 0.1 K. The pointing accuracy of the telescope was within $\sim 20''$. The raw data were reduced with the DOS-version of FCRAO SPA software as revised by the TRAO. The baseline of each spectrum was fitted in two steps: After a baseline was corrected with the best fit of a polynomial up to 3rd order, its remaining sinusoidal baseline component was further removed with a sine fitting. The resulting line profile was fitted into a Gaussian to obtain line parameters such as the $\Delta V$s (FWHMs). The data are presented in the form of maps made using the NRAO AIPS software package.

## 3. THE DATA

### 3.1 Total Velocity Integrated Map

The intensity distribution integrated over the entire velocity range of the molecular clouds is shown in Figure 3. This shows a global distribution of the GC molecular clouds. The entire morphology is fairly similar to those seen at 110 GHz $^{13}$CO and 98 GHz CS surveys by Bally *et al.* (1987, 1988). A noticeable feature is the asymmetric distribution in the integrated emission, *i.e.*, the emission has a lopsided distribution concentrated in the first quadrant ($l > 0$). The GC molecular clouds on the map are clustered into four large complexes that are $l \approx 5.5°$ complex, $l \approx 3.2°$ complex (Bania Clump 2), Galactic nuclear disk complex (within $-2° \leq l \leq 2°$), and $l \approx -5.0°$ complex (Bania Clump 1). Kinematics and physical properties of these complexes are different from each other. Detailed features of these groups will be described in forthcoming paper.

### 3.2 The Velocity Channel Maps

Two sets of channel maps are made: one binned into 100 km s$^{-1}$ intervals (Figure 4) and the other binned into 20 km s$^{-1}$ intervals (Figure 5). From a visual inspection, these maps show that most of the clouds follow the velocities permitted by the 'regular' Galactic rotation. 'Regular' means that the Galaxy rotates in the sense that clouds on the first quadrant ($l > 0$ on the map) recede ($V_{LSR} > 0$) whereas those on the fourth quadrant ($l < 0$) approach us ($V_{LSR} < 0$). One outstanding feature, however, is that the $V_{LSR} > 0$ components in the first quadrant are brighter and much more abundant than those in the fourth quadrant. Our channel maps also show weak emission components that are "forbidden" by Galactic rotation, as opposed to 'regular'. These may be the components moving on a non-circular orbit due to a non-axisymmetric potential in the GC. It should be emphasized that there are numerous clumps in the Galactic nuclear disk complex (within $-2° \leq l \leq 2°$) that have forbidden velocities. Bania Clump 2 also has forbidden components at $-20 \sim 0$ km s$^{-1}$. Bania Clump 1 (60$\sim$120 km s$^{-1}$) also has several components which are really highly forbidden.

It should be noted that channel maps that are binned into 100 km $s^{-1}$ interval are very similar to those observed at $^{13}CO$ by Bally *et al.* (1988) where emission components in velocity interval between $-20 \sim 20$ km $s^{-1}$ have been excluded to avoid contamination due to the disk gas emission. Our maps in fact are free from any Galactic diffuse gas component around $V_{LSR} = 0$ km $s^{-1}$. This is because HCN molecule is collisionally excited in regions of $n \geq 10^5$ cm$^{-3}$ (Turner 1974) and such gas is rare in the outer galactic disk. Unlike the $^{13}CO$ case, since the $-20 \sim 20$ km $s^{-1}$ component need not be subtracted from the HCN maps and structures between -20 and 20 km $s^{-1}$ are all preserved. These structures are shown in Figure 5, where the distribution of the components within $|l| \geq 2°$ are fairly symmetrical around to $l = 0$, due to the existence of comparable amount of forbidden velocity material.

### 3.3 Space - Velocity Maps

To show the velocity structures of the molecular clouds, various space-velocity maps are presented. The averaged global velocity structures along the $l$ or $b$ directions are shown by the $l - v$ diagram (averaged over $b$) in Figure 6 and the $b - v$ diagram (averaged over $l$) in Figure 7. Structures of molecular clouds along any $l$ or $b$ directions are also shown in the $l - v$ maps at any $b$ in Figure 8 and the $b - v$ maps at any $l$ in Figure 9.

#### 3.3.1 The $l - v$ Diagram Averaged Over $b$

Figure 6 shows the global velocity structures averaged over $b$ as a function of $l$. The global emission features are similar to those seen in $^{13}CO$ by Bally *et al.* (1988), except that the degree of asymmetry in the emission is much greater than that seen in $^{13}CO$; the emissions are heavily lopsided ($\geq 80\%$ of overall emissions) on the side of the first quadrant and $V_{LSR} > 0$. These emissions extend to $l \approx -0.3°$, from which the emissions of $V_{LSR} < 0$ become dominant at the fourth quadrant. But this does not seem to mean that the center of the rotation is at $l \approx -0.3°$ because the components around $l \approx -0.2°$ (Sgr A clouds) can be considered as those plunging across the GC region (Brown & Liszt 1984) or in non-circular motion.

The 300 pc Expanding Molecular Ring is only barely seen in HCN. However, like the channel maps in §3 (b), the $l-v$ map does not show any diffuse disk gas emissions between $-20 \sim 20$ km s$^{-1}$. Moreover, the 3 kpc arm component clearly seen on the $^{13}$CO map is not visible on the $l-v$ map. It is therefore possible to see the GC molecular clouds through the HCN line without contamination from foreground components.

### 3.3.2 The $b-v$ Diagram Averaged Over $l$

This diagram shows that most emissions are located within $-0.4° \leq b \leq 0.4°$ and their distribution in velocity is extremely asymmetric: most emission is at $V_{LSR} > 0$. But the distribution is nearly symmetric to $b = -0.0667°$, with the emission peak at $V_{LSR} \approx 50$ km s$^{-1}$. Noticeable is a slope in the map from the lower left to the upper right direction: the velocity tends to increase with b. This can also be seen in the $l-v$ diagrams (Figure 8): the $V_{LSR} < 0$ components stem mainly from the low latitude components (e.g. Sgr C, E) located in the fourth quadrant whereas the $V_{LSR} > 0$ components stem mainly from the high latitude components at the first quadrant.

### 3.4 Distribution of $\Delta V$ (FWHM)

We have attempted to decompose all the line profiles into Gaussian components. In most cases, the line profiles are well separated with different $V_{LSR}$. An example of the decomposition is shown in Figure 10. We have obtained $\Delta V$ and $V_{LSR}$ for 886 components from the decomposition. The distribution of number of components in the interval of $\Delta V \sim \Delta V + \delta(\Delta V)$ (=10 km s$^{-1}$) is shown in Figure 11. As expected from the fact that the HCN clouds are relatively dense and that many clouds can be overlapped within the beam of the telescope, most of the line profiles are wide. About 80 % of the $\Delta V$s lie in the range between 15–55 km s$^{-1}$ with the mode at $\Delta V \approx 25$ km s$^{-1}$. There are even clouds with $\Delta V \geq 100$ km s$^{-1}$. If we impose the constraint of $\bar{\rho} > \rho_{tid}$ ($\approx 5 \times 10^{-21}$ g cm$^{-3}$), where $\bar{\rho}$ is a mean density of a cloud and $\rho_{tid}$ is the density to overcome the Galactic tidal field,

$$\Delta V \sim G^{1/2}\,\rho_{tid}^{1/6}\,M^{1/3} \approx 13.5\ (\mathrm{km\ s^{-1}})(\frac{\rho_{tid}}{5\times10^{-21}\mathrm{g\ cm^{-3}}})^{1/6}\,(\frac{M}{10^{6}\,\mathrm{M_{\odot}}})^{1/3}.$$

Therefore the typical value of $\Delta V \approx 25$ km s$^{-1}$ seen in the Figure 11 may imply that the typical mass of most HCN clouds is of order $10^6$ M$_\odot$. The much larger value of $\Delta V \approx 100$ km s$^{-1}$ appears to be impossible for a single cloud, however. This upper value may instead indicate the velocity dispersion for several clouds with slightly different $V_{LSR}$ within a single beam of the telescope. This is fairly probable given that the velocity dispersion of stars in the Galactic Center region is in fact ~100 km s$^{-1}$ (Sellgren *et al.* 1990; Blum *et al.* 1994).

Obtained values of all $\Delta V$s are plotted as a function of $l$ in Figure 12. Large $\Delta V$s are seen toward the directions showing strong HCN emission. In positive $l$, the maximum value of $\Delta V$ tends to decrease with $l$ while such a tendency is seen only weakly in the region of negative $l$ because of the scarcity of HCN clouds. The dependence of the maximum $\Delta V$ on $l$ may be partly due to the overlapping effects of clouds.

## 4. SUMMARY

The regions in the GC were surveyed with HCN J=1–0 line over a wider regions ($-6^\circ \leq l \leq 6^\circ$ and $-0.8^\circ \leq b \leq 0.87^\circ$) with a better or comparable grid resolution than previous molecular line surveys. Two unique characteristics of this survey are noteworthy. First, HCN J=1–0 line is optically thin ($\tau \leq 0.3$) and therefore the data give apparently unsaturated pictures of the molecular clouds. Secondly, since it is a good tracer for high density regions (n$\geq 10^5$ cm$^{-3}$), less contamination by either foreground or background components was seen and thus finer structures were observable. Conclusions reached with the present observations are summarized as follows.

The HCN J=1−0 maps show that the most of dense clouds follow the velocities permitted by the 'regular' Galactic rotation. One outstanding feature, however, is that the overall distribution is heavily lopsided in that more than 80% of all emissions reside in the first quadrant ($l > 0$) and $V_{LSR} > 0$. The degree of the asymmetry exceeds greatly that found in the $^{13}$CO map by Bally *et al.* (1988). The HCN maps also confirm that there are several forbidden velocity components which may be due to highly non-circular motions.

The most of the detected lines were found to have wide $\Delta V$ of 15$\sim$55 km s$^{-1}$ with the mode at $\Delta V \approx 25$ km s$^{-1}$. The typical value of $\Delta V \approx 25$ km s$^{-1}$ may imply that mass of most HCN clouds in the GC region is an order of $10^6\,M_{\odot}$. It is also shown that the maximum $\Delta V$ tends to decrease with $l$, which may be partly due to the overlapping effects of clouds.

I am most greatful to Prof. H. M. Lee for his heartful guidance and supervision, and to Prof. H. B. Ann for sincere advice and assistance. I would like to thank Prof. K.-T. Kim, Dr. Y. C. Minh and Y. Lee for their much critical discussions and useful advice. Several helpful discussions with Dr. S.H. Cho, J. H. Jung and Y. S. Park are also greatfully acknowledged. My sincere thanks are also given to the staff of TRAO, who are Dr. C. H. Lee, S. T. Han, H. S. Chung, H. G. Kim, H. R. Kim, I. S. Yim and T. S. Kim. It would have been impossible to complete this survey without their help. This paper was finally polished by an anonymous refree, to whom I would like to express my thanks. This research was supported in part by the grant for observers of Korean Astronomical Obervatory.

Table 1. Previous molecular line surveys toward the Galactic Center

| Molecule | Freq.(GHz) | Longitude | Latitude | Grid ($l \times b$) | Beam | Ref. |
|---|---|---|---|---|---|---|
| OH | 1.665 | $-6.0^\circ \le l \le 8.6^\circ$ | $-1.0^\circ \le b \le 1.0^\circ$ | $10' \times 12'$ | $3.0'$ | (1) |
| $^{12}$CO | 115.271 | $-5.0^\circ \le l \le 122^\circ$ | $-1.0^\circ \le b \le 1.0^\circ$ | $3'$ | $1.7'$ | (2) |
| $^{13}$CO | 110.201 | $-2.0^\circ \le l \le 2.0^\circ$ | $-0.5^\circ \le b \le 0.5^\circ$ | $4' \times 8'$ | $1.7'$ | (3) |
| $^{13}$CO | 110.201 | $-5.0^\circ \le l \le 5.0^\circ$ | $-0.6^\circ \le b \le 0.6^\circ$ | $6'$ | $1.7'$ | (4) |
| CS | 97.981 | $-1.0^\circ \le l \le 3.7^\circ$ | $-0.4^\circ \le b \le 0.4^\circ$ | $2'$ | $1.9'$ | (4) |
| $H_2CO$ | 4.830 | $0.5^\circ \le l \le 4.0^\circ$ | $-0.5^\circ \le b \le 0.9^\circ$ | $6'$ | $2.9'$ | (5) |

References. — (1) Cohen & Dent (1983); (2) Stark *et al.* (1988); (3) Heiligman (1987); (4) Bally *et al.* (1987); (5) Zylka *et al.* (1992).

---

## FIGURE CAPTIONS

**Fig. 1** Surveyed region. Observed positions are marked in dots whose size approximately corresponds to the HPBW (61 arcsec) of the Telescope.

**Fig. 2** Sample spectra of $H^{12}CN$ J=1−0 lines detected at (a) $(l, b) = (5.4667°, -0.3333°)$, (b) $(l, b) = (3.2667°, 0.4667°)$, (c) $(l, b) = (1.4667°, 0°)$, (d) $(l, b) = (0.8°, -0.0667°)$, (e) $(l, b) = (0°, -0.0667°)$, (f) $(l, b) = (-0.5333°, -0.1334°)$, (g) $(l, b) = (-1.0°, -0.3333°)$, and (h) $(l, b) = (-5.4°, 0.3333°)$. The spectrum of (h) is a representative example having components forbidden by Galactic rotation.

**Fig. 3** Total velocity integrated intensity map. Integrated Intensity in this map is $\int T_A^* dv/\delta v$ where $\delta v$ is channel width (3.38 km $s^{-1}$). It is shown that most of emissions are asymmetrically distributed at the the first quadrant ($l > 0$).

**Fig. 4** Velocity channel maps binned at 100 km $s^{-1}$ intervals. Integrated intensity in this map is $\int T_A^* dv/\delta v$ where $\delta v$ is channel width (3.38 km $s^{-1}$). Most of the emissions are distributed in the positive velocity maps, while some channel maps contain several emission features forbidden by Galactic rotation which may be in non-circular orbit.

**Fig. 5** Velocity channel maps binned into 20 km $s^{-1}$ intervals. Integrated Intensity in this map is $\int T_A^* dv/\delta v$ where $\delta v$ is channel width (3.38 km $s^{-1}$). The scale of b coordinate in the maps is stretched a fator of two for viewing convenience.

**Fig. 6** Galactic longitude - velocity diagram averaged over $b$. Most emission ($\geq$ 80% of all emission) is found in the first quadrant side ($l > 0$) and $V_{LSR} > 0$. In this map foreground or background emissions such as diffuse outer galactic disk gas components and 3 kpc arm components seen in CO maps are not seen.

**Fig. 7** Galactic latitude - velocity diagram averaged over $l$.

**Fig. 8** Galactic longitude - velocity diagram at any $b$ within $-0.6° \leq b \leq 0.67°$.

**Fig. 9** Galactic latitude - velocity diagram at any $l$. In this diagram maps of strongly emitting region are selectively presented.

**Fig. 10** Gaussian fitting example of a line detected at the position of $(l, b) = (0.4667°, -0.0667°)$.

**Fig. 11** $N_{bin}/N_{tot}$ versus the $\Delta V$. $N_{tot}$ is a total number (886) of the decomposed Gaussian components and $N_{bin}$ is a number of Gaussian components in the interval $\Delta V \sim \Delta V + \delta(\Delta V)(= 10\text{km s}^{-1})$. The mode is at $\Delta V \approx 25$ km s$^{-1}$.

**Fig. 12** $\Delta V$ variation along $l$. The maximum $\Delta V$ tends to decrease with $l$.

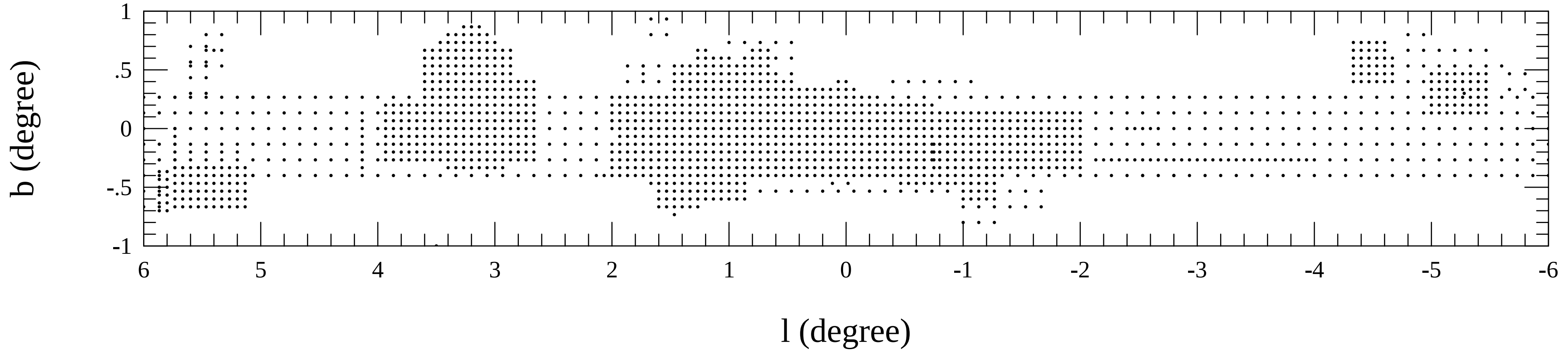

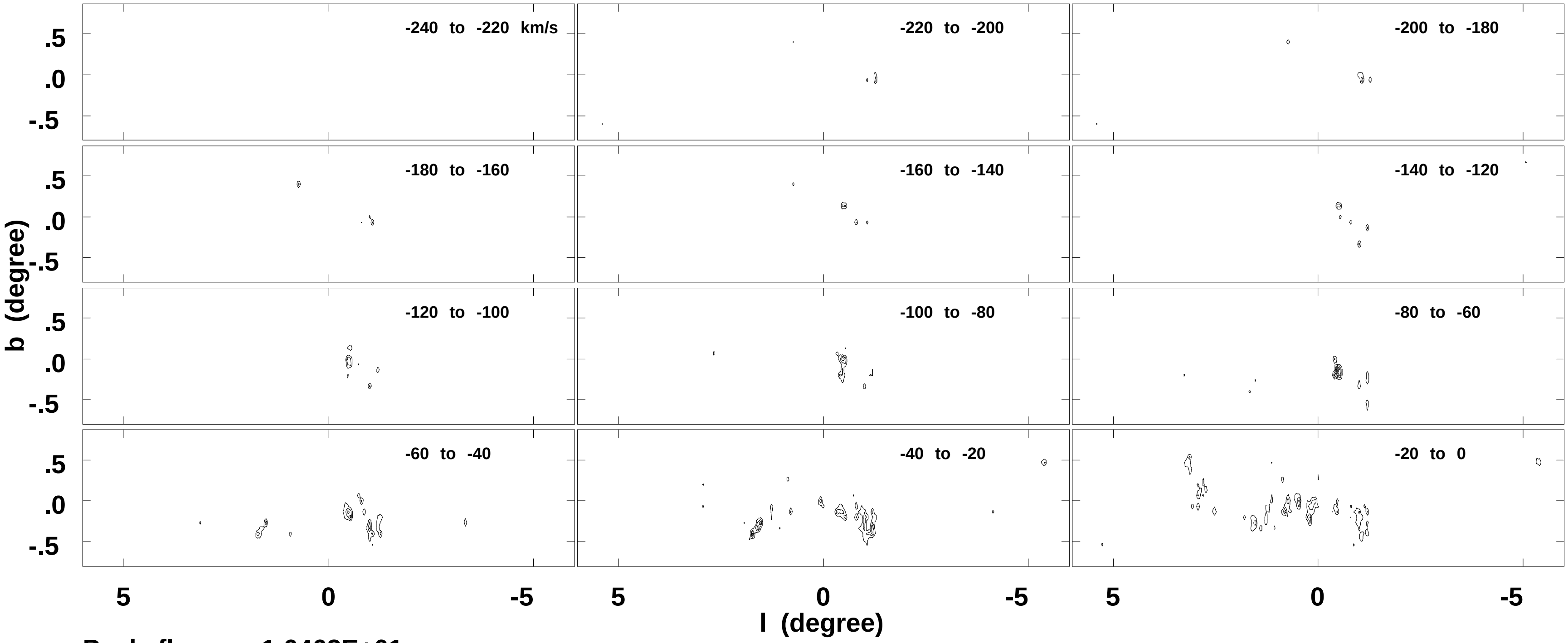

Peak flux = 1.6462E+01
Levs = 1.0000E+00 * ( 1.000, 2.000, 3.000, 4.000, 5.000, 6.000, 7.000, 8.000, 9.000, 10.00, 11.00, 12.00, 13.00, 14.00, 15.00, 16.00)

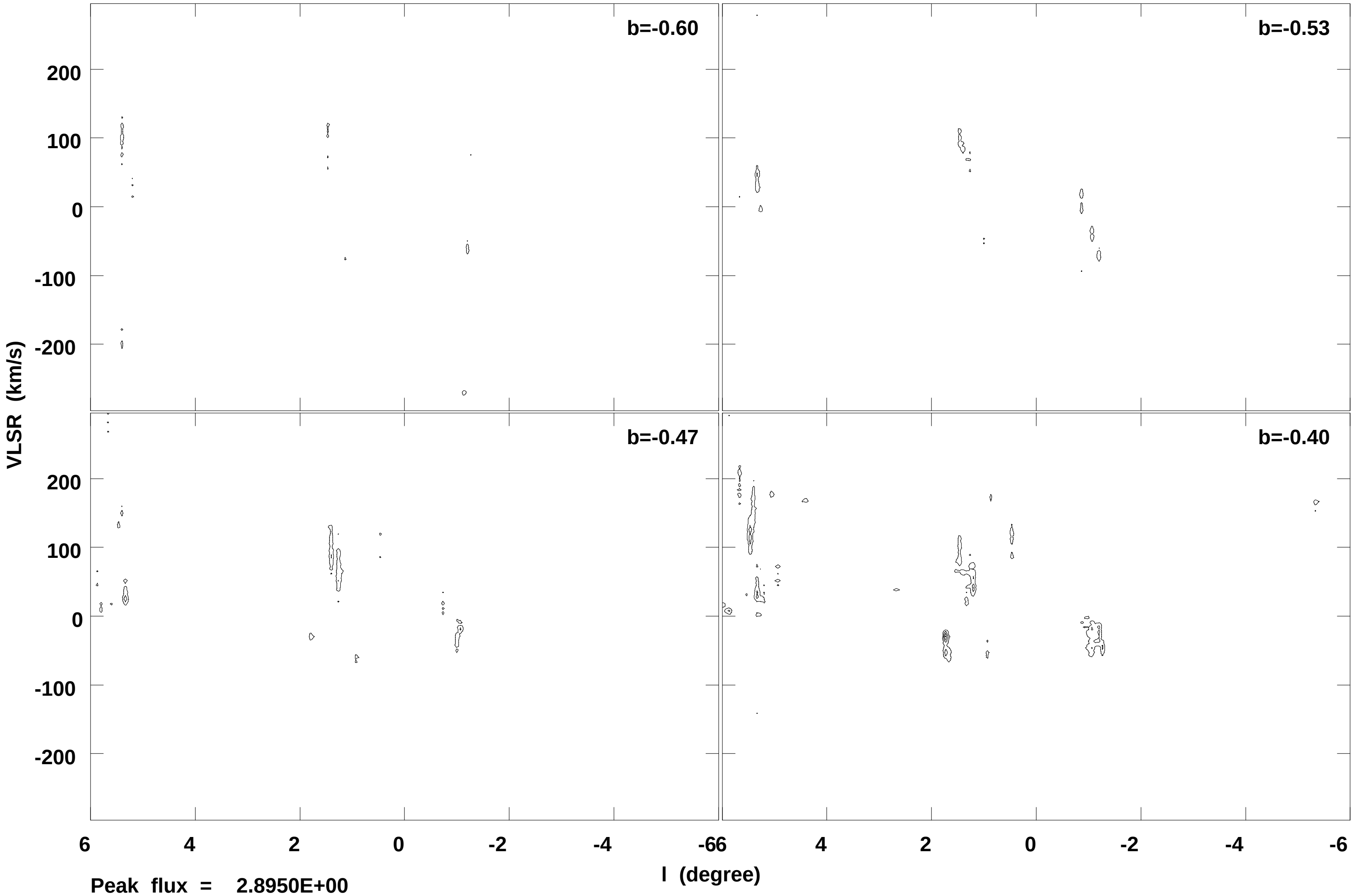

Peak flux = 2.8950E+00
Levs = 1.8000E-01 * ( 1.000, 2.500, 4.000,
5.500, 7.000, 8.500, 10.00, 11.50, 13.00,
14.50, 16.00)

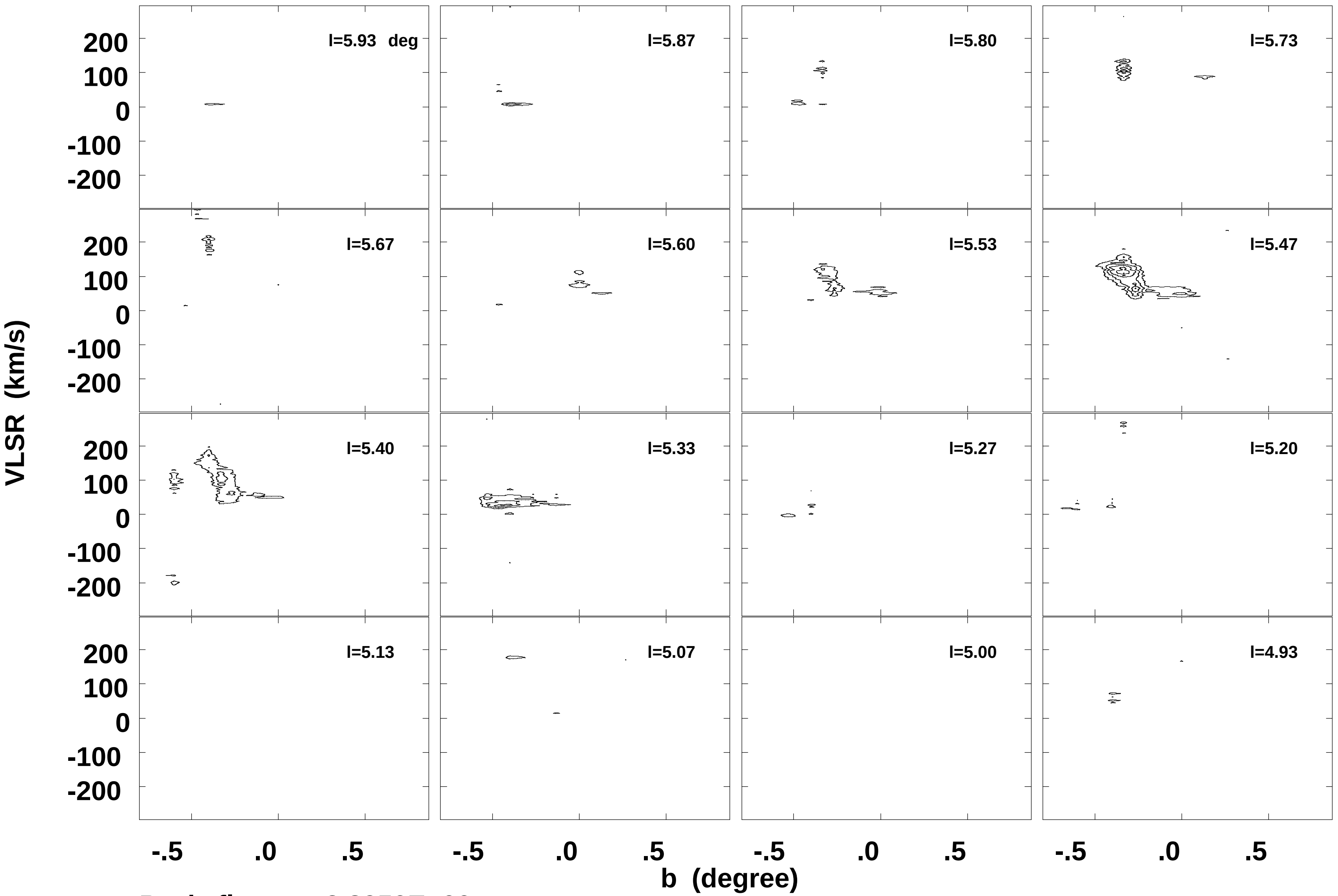

Peak flux = 2.8950E+00
Levs = 1.8000E-01 * ( 1.000, 2.000, 3.000, 4.000, 5.000, 6.000, 7.000, 8.000, 9.000, 10.00, 11.00, 12.00, 13.00, 14.00, 15.00, 16.00)

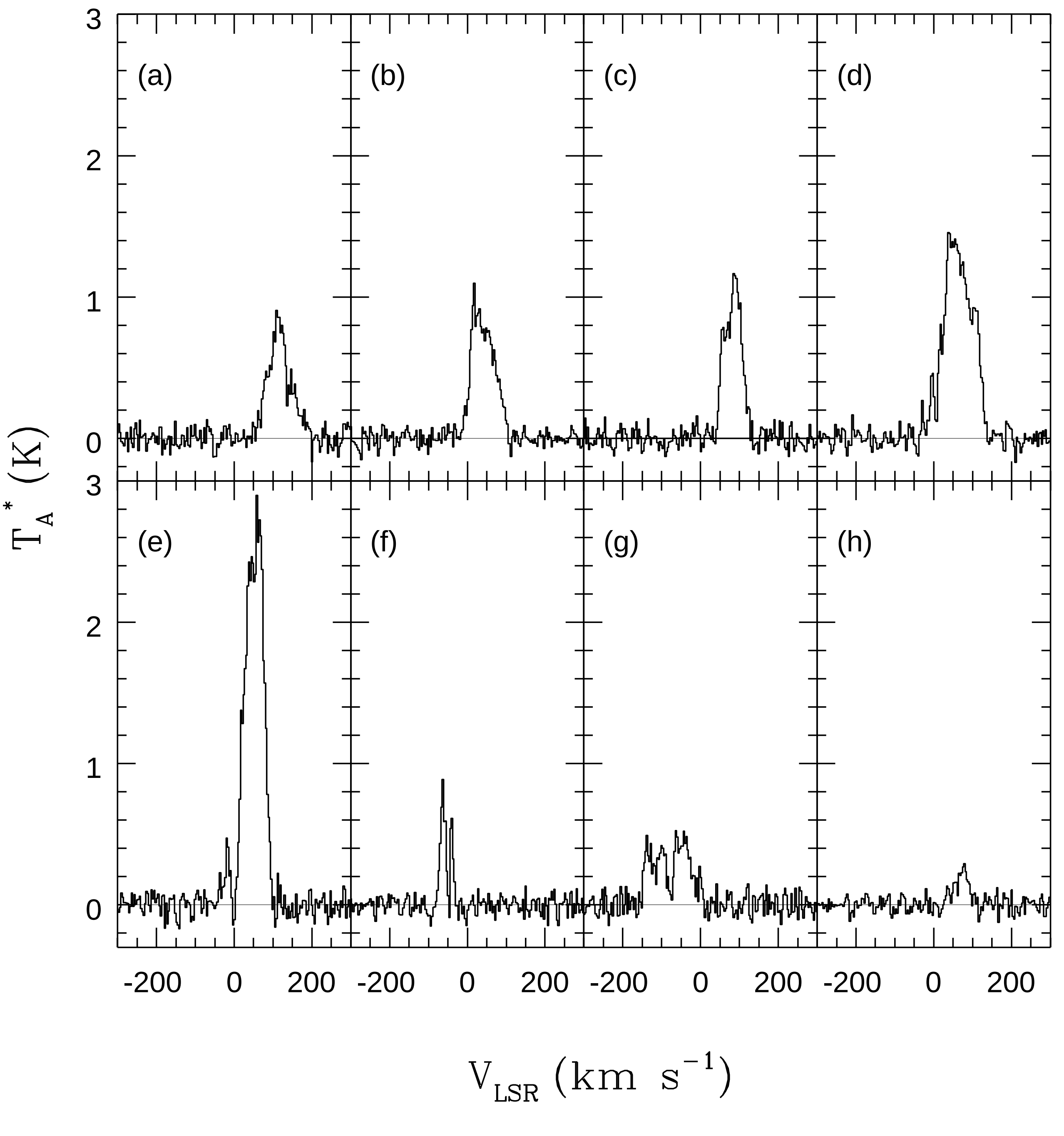

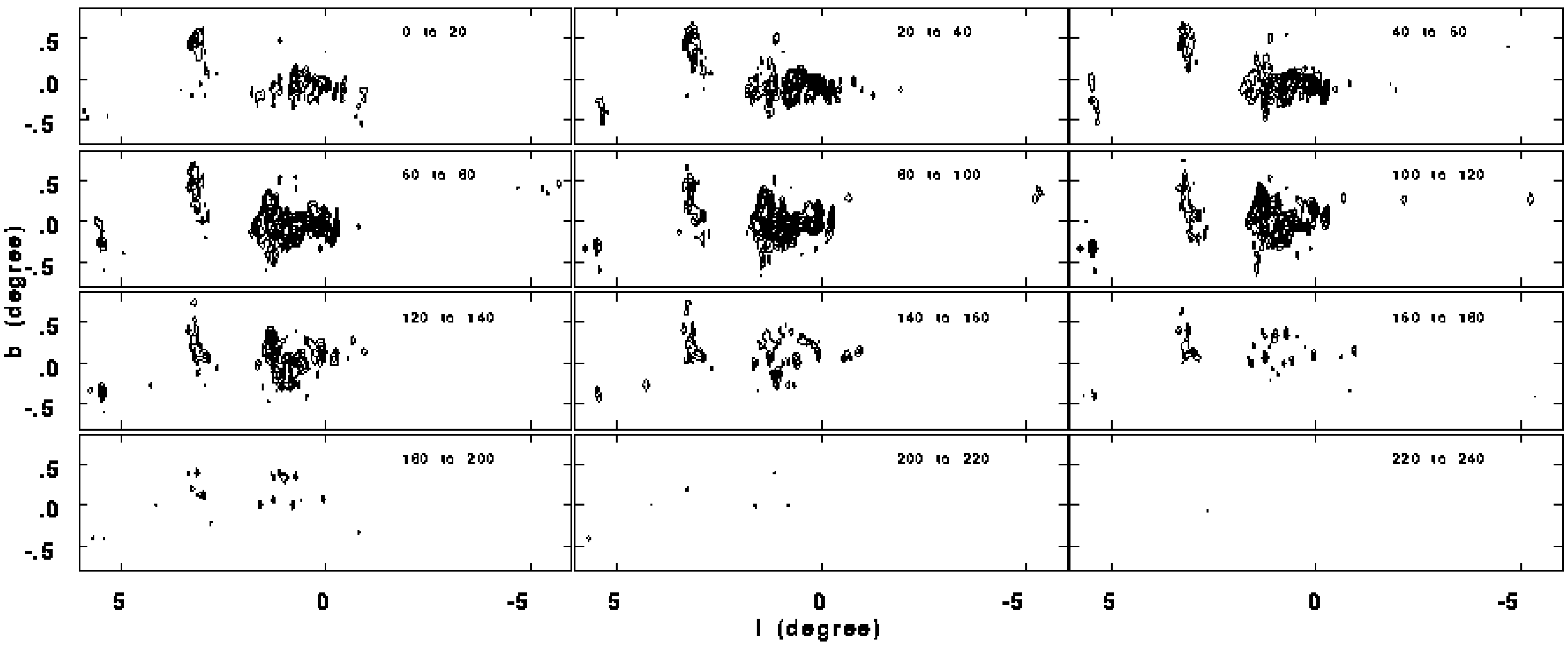
0 to 20
20 to 40
40 to 60
60 to 80
80 to 100
100 to 120
120 to 140
140 to 160
160 to 180
180 to 200
200 to 220
220 to 240
b (degree)
l (degree)

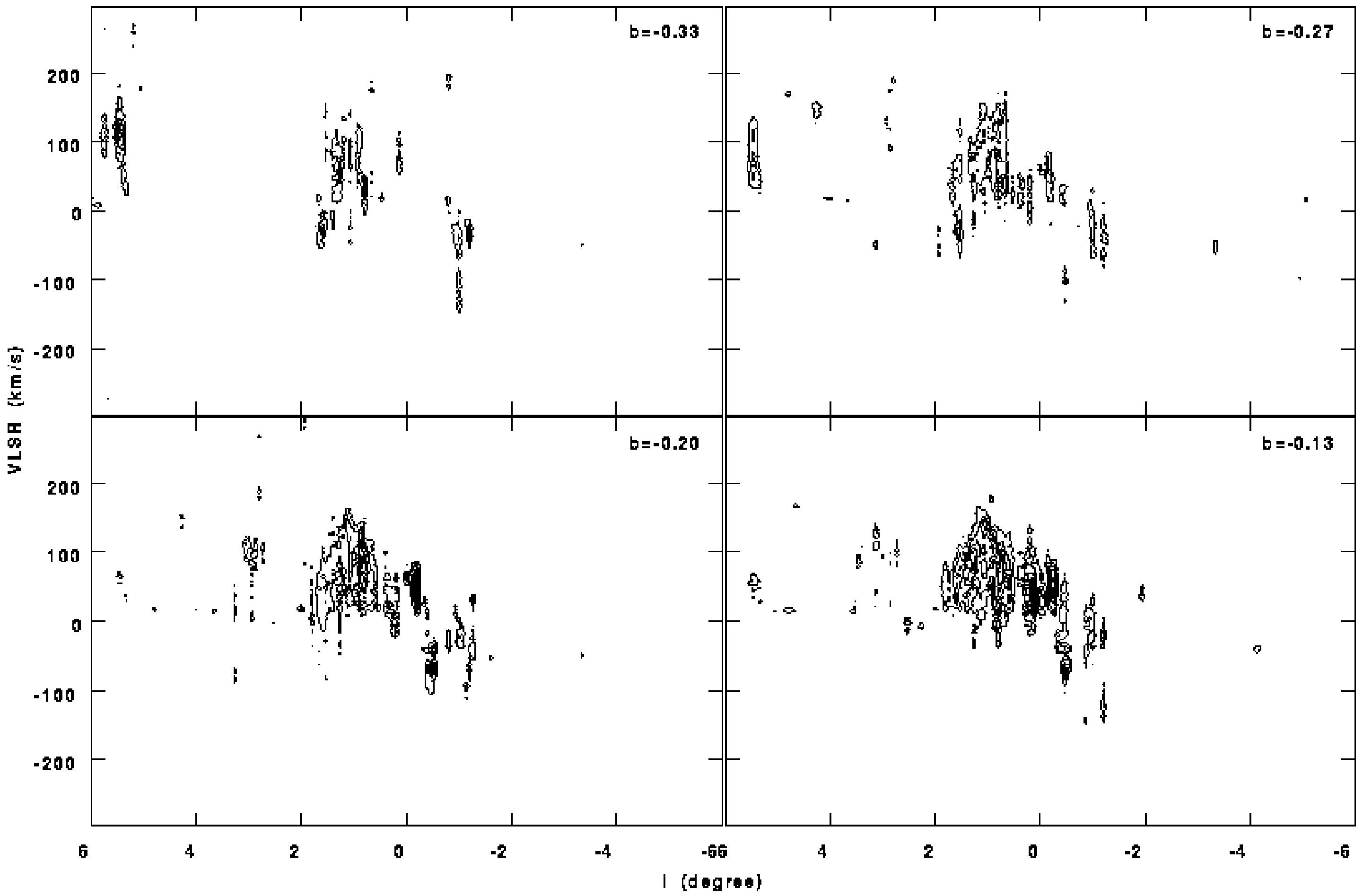

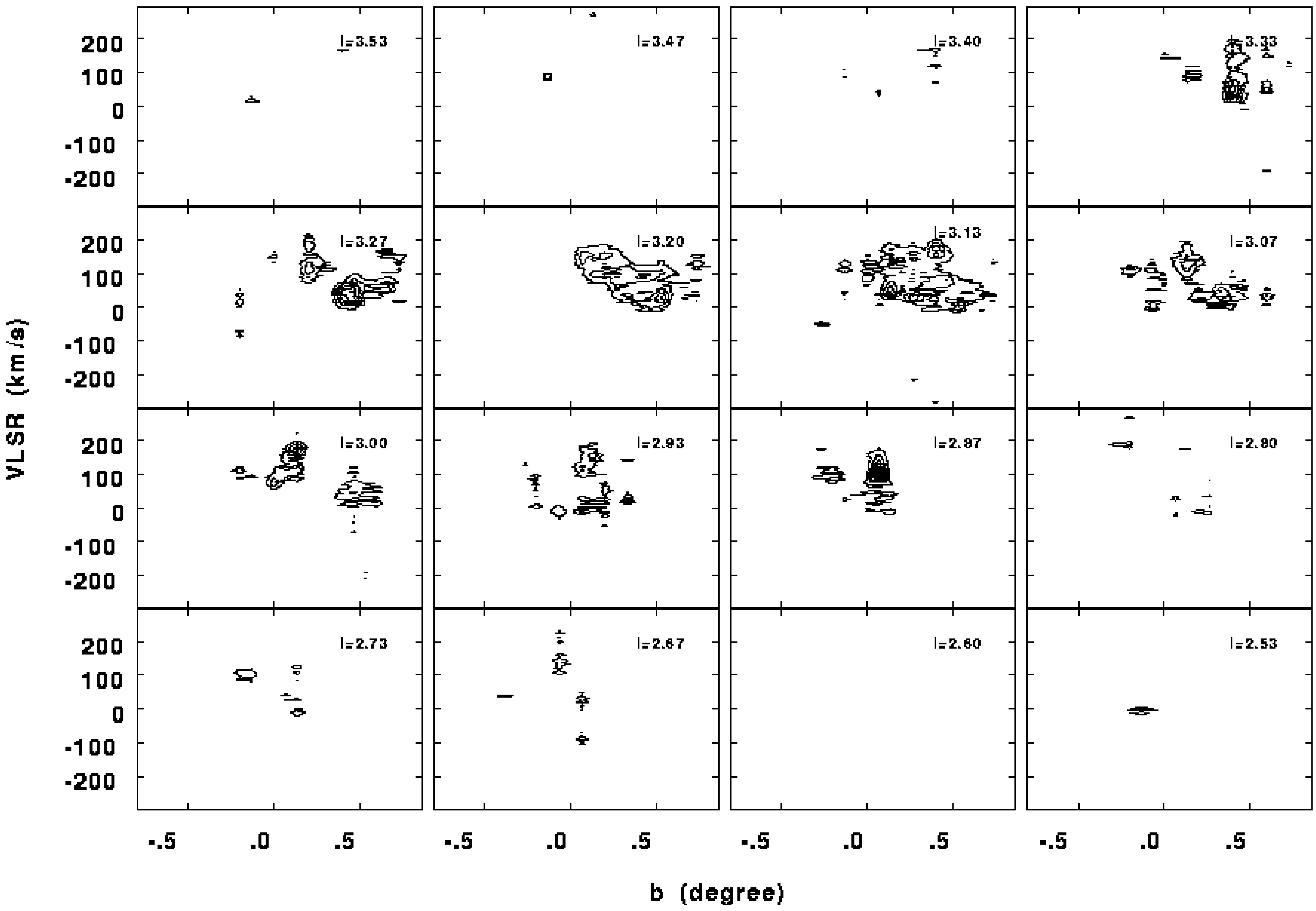

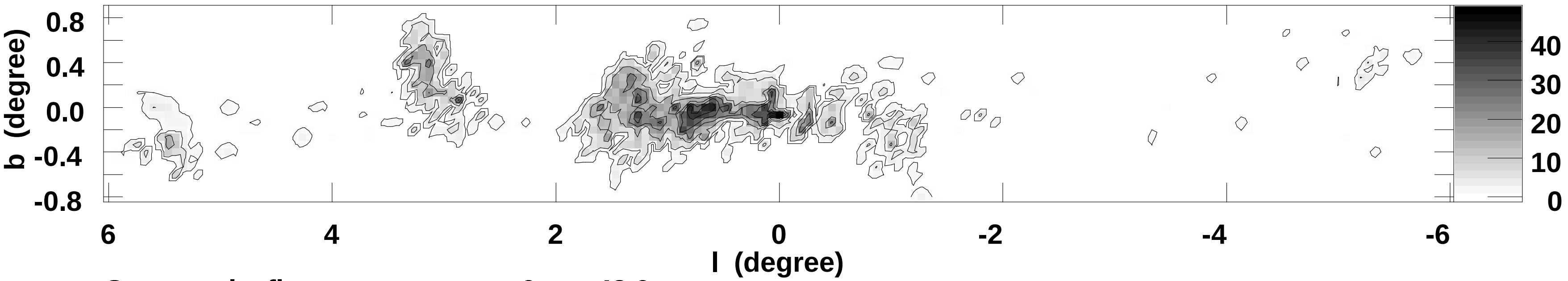

Grey scale flux range= .0 48.0
Peak contour flux = 4.7415E+01
Levs = 4.0000E+00 * ( .250, 1.000, 3.000,
5.000, 7.000, 9.000, 11.00)

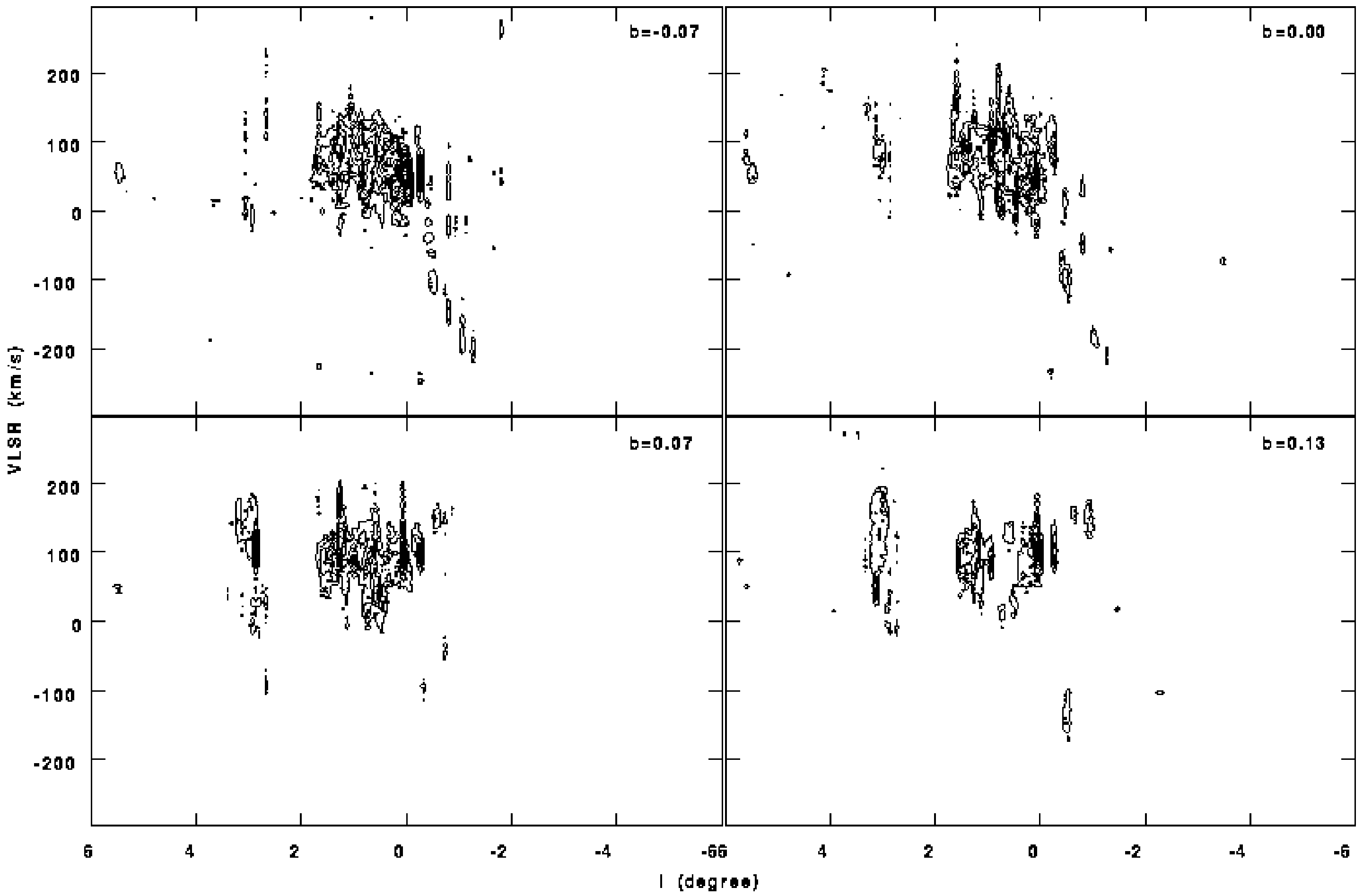

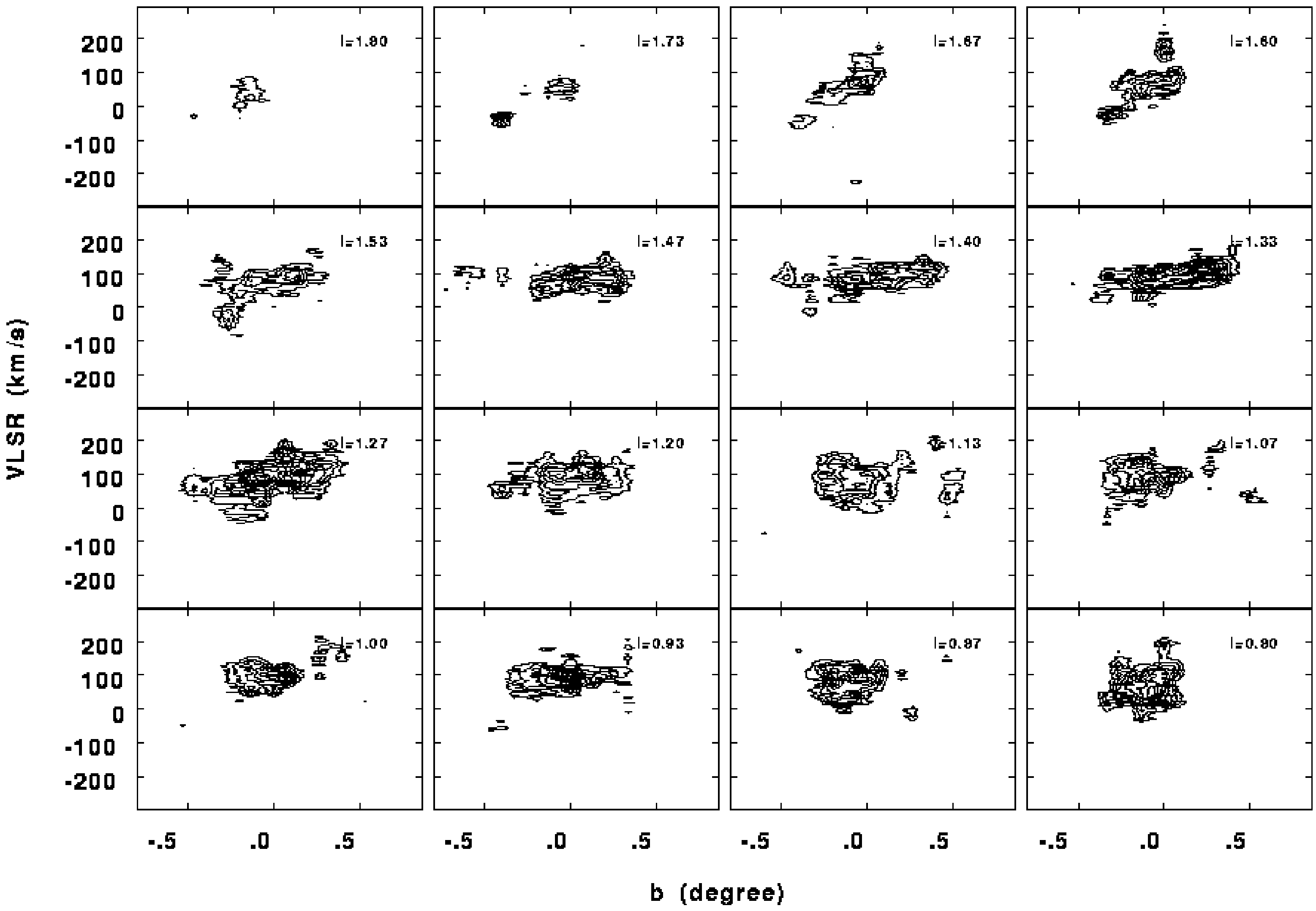

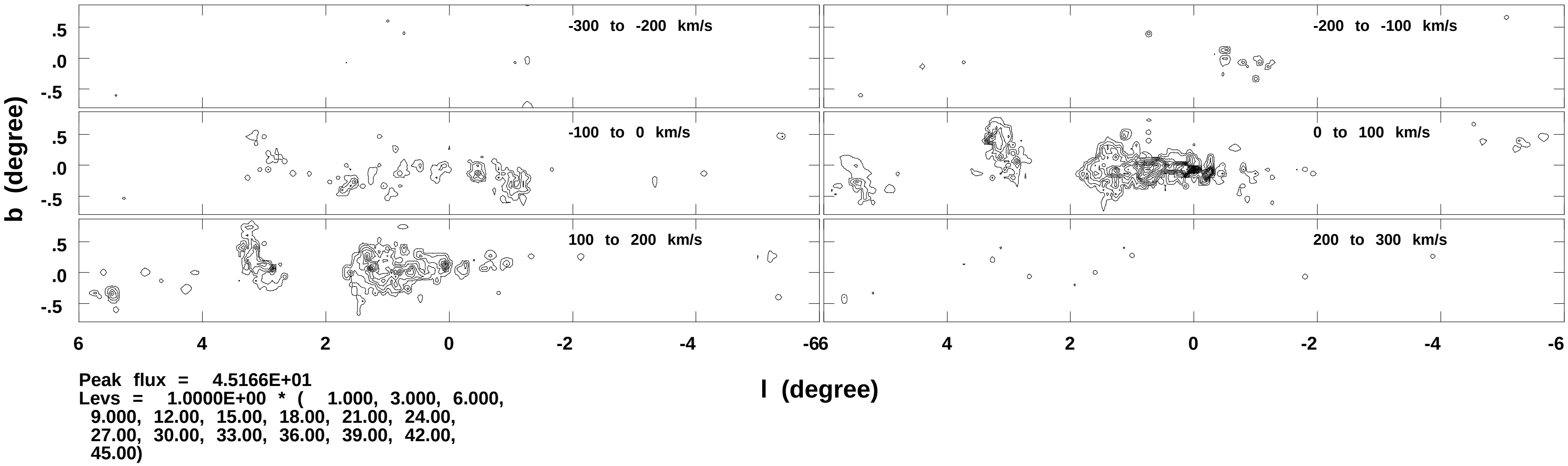

-300 to -200 km/s
-200 to -100 km/s
-100 to 0 km/s
0 to 100 km/s
100 to 200 km/s
200 to 300 km/s
b (degree)
.5
.0
-.5
6
4
2
0
-2
-4
-66
-6
l (degree)
Peak flux = 4.5166E+01
Levs = 1.0000E+00 * ( 1.000, 3.000, 6.000,
9.000, 12.00, 15.00, 18.00, 21.00, 24.00,
27.00, 30.00, 33.00, 36.00, 39.00, 42.00,
45.00)

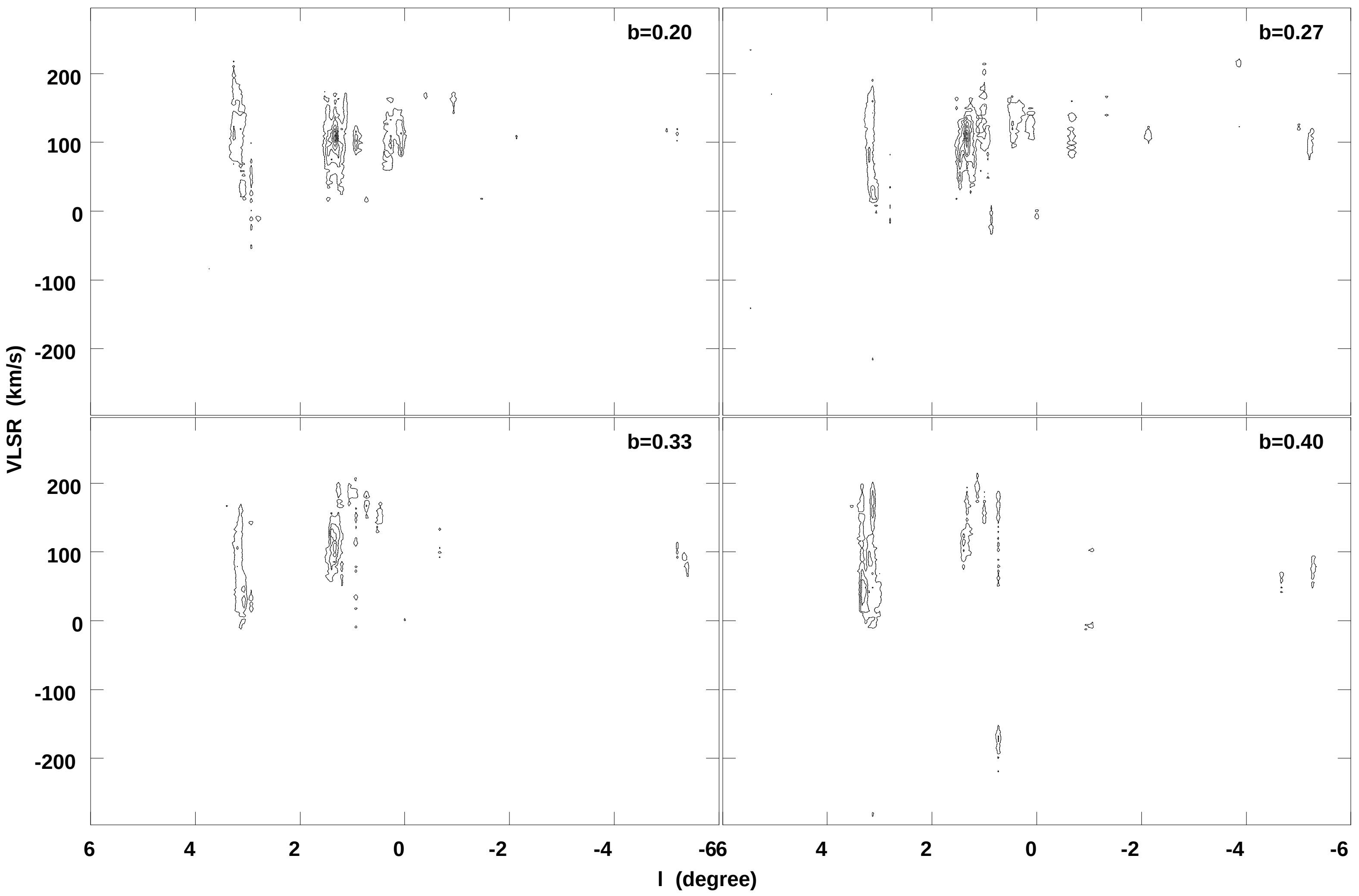

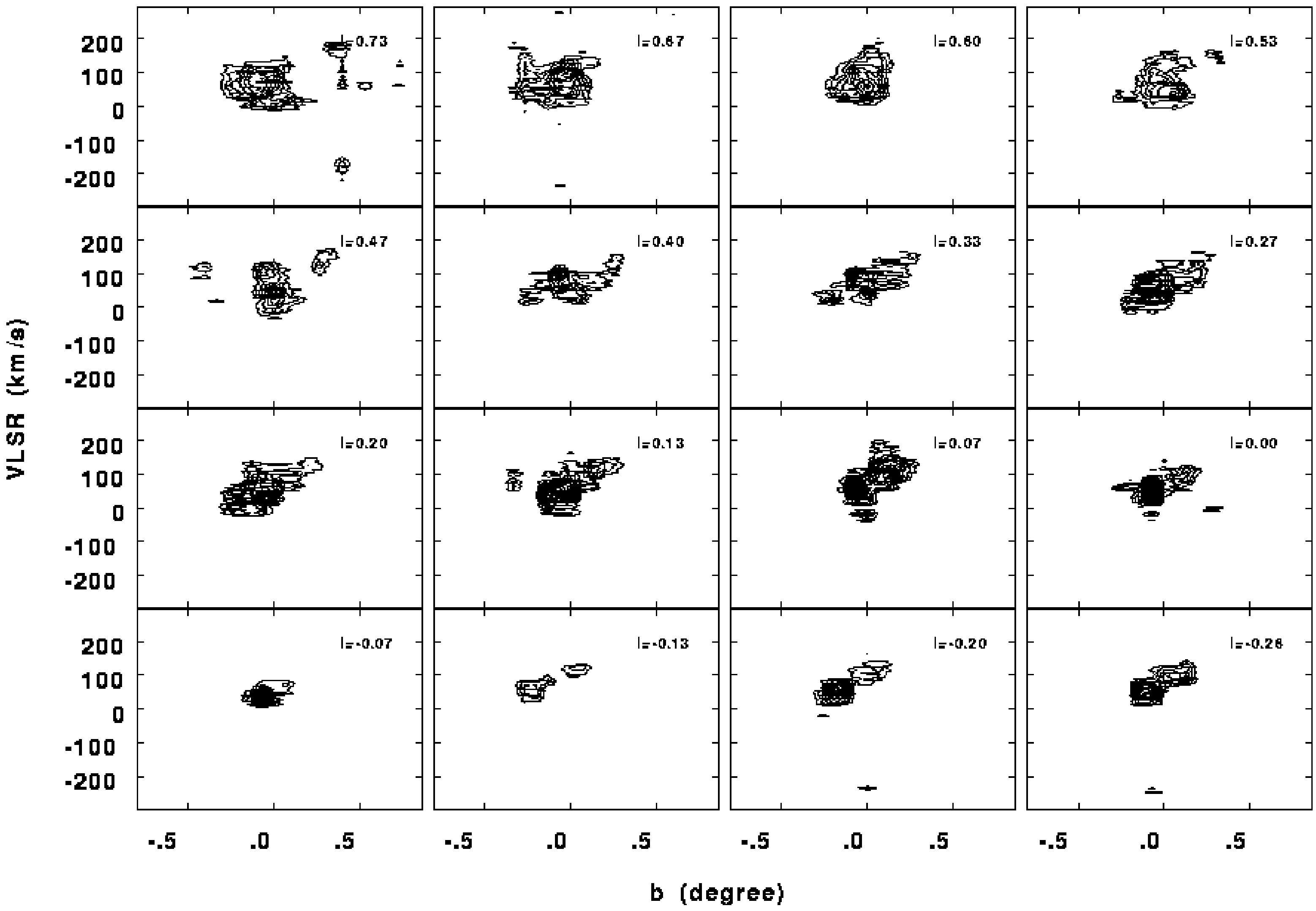

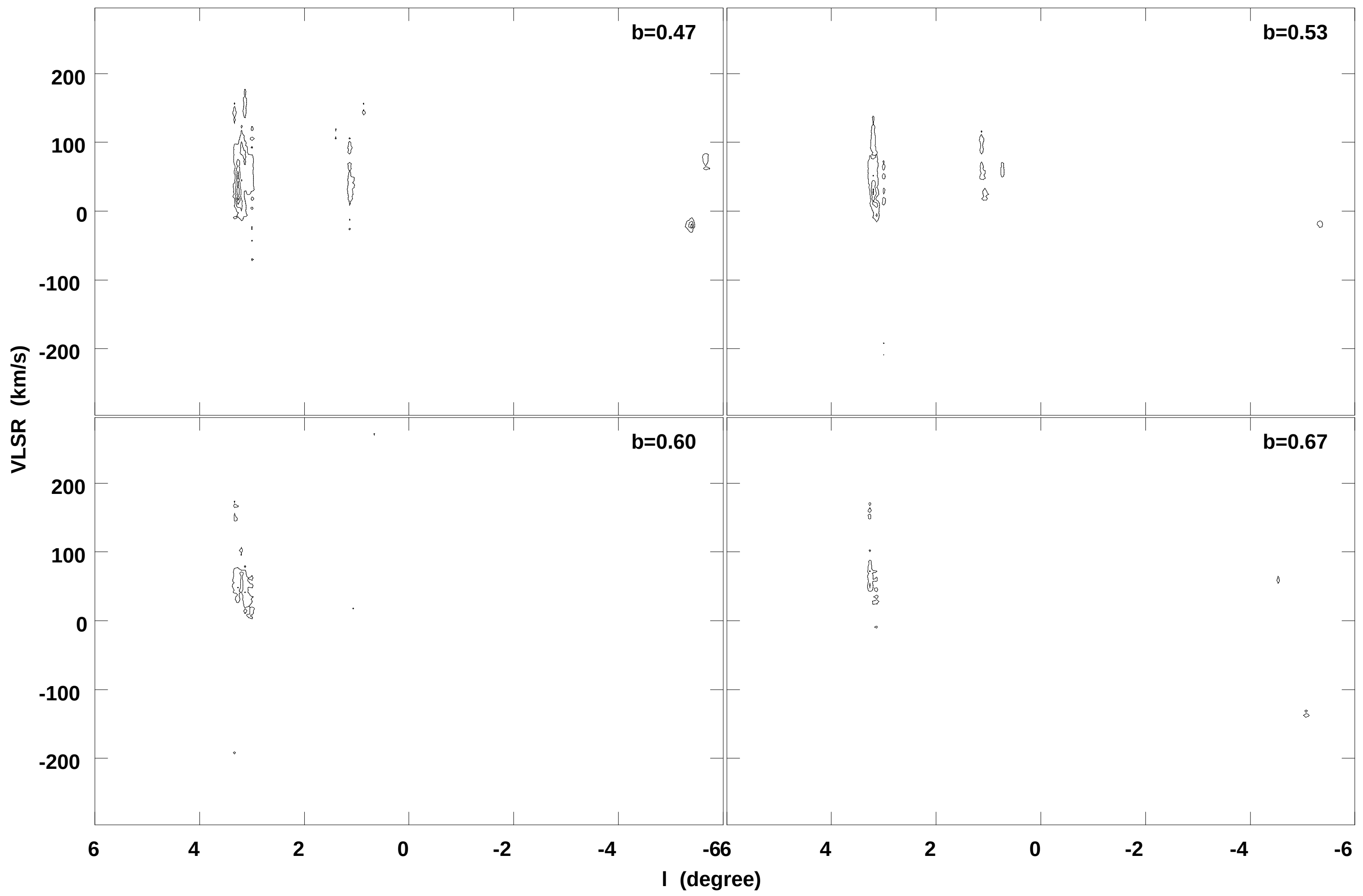

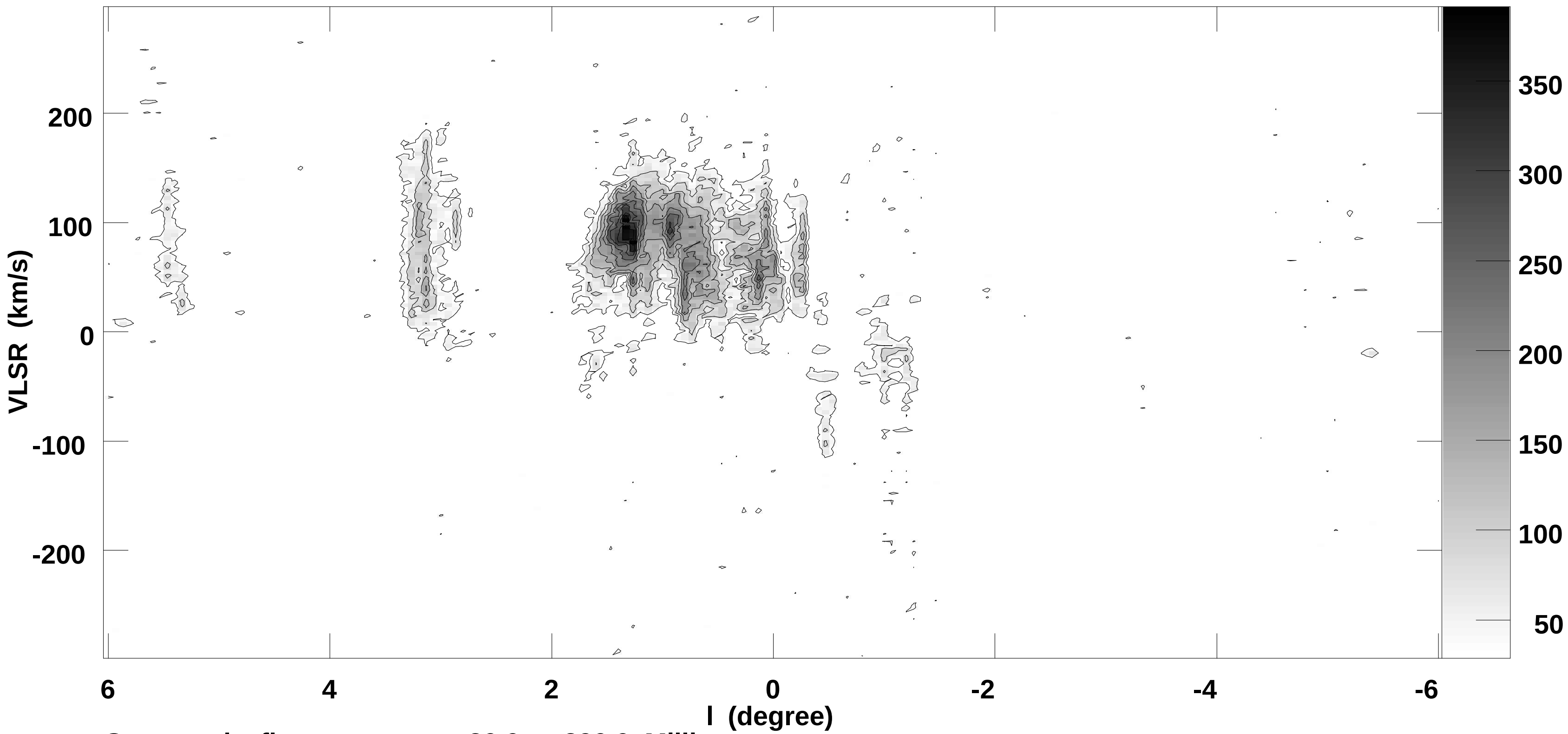

Grey scale flux range= 30.0 390.0 Milli
Peak contour flux = 3.8450E-01
Levs = 3.8450E-02 * ( .900, 2.000, 3.000,
4.000, 5.000, 6.000, 7.000, 8.000, 9.000,
10.00)

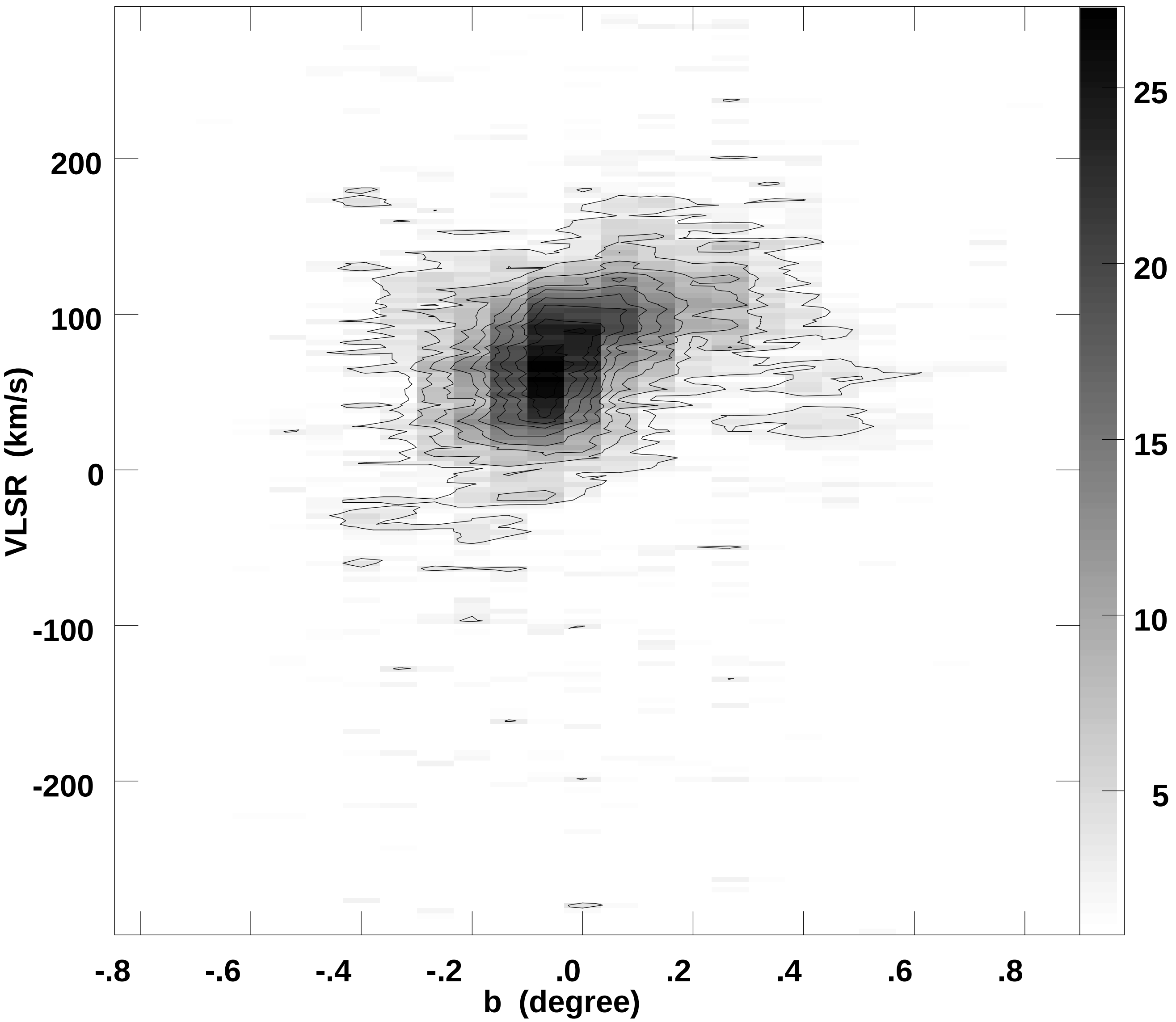

Grey scale flux range= 1.0 27.2
Peak contour flux = 2.7219E+01
Levs = 2.7219E+00 * ( 1.000, 2.000, 3.000,
4.000, 5.000, 6.000, 7.000, 8.000, 9.000,
10.00)